\documentclass{iau}
\usepackage{natbib}
\usepackage{amsmath}
\usepackage{graphicx}
\usepackage{multirow}
\usepackage{orcidlink}
\usepackage{aas_macros}

\begin{document}

\lefttitle{Padois et al.}
\righttitle{Constraining exoplanet population parameters with {\it Kepler}}

\jnlPage{1}{7}
\jnlDoiYr{2026}
\doival{10.1017/xxxxx}

\aopheadtitle{Proceedings IAU Symposium 408}
\editors{G. Buldgen, A. Vidotto, \& A. Miglio, eds.}

\title{Constraining exoplanet population parameters \\ with {\it Kepler} and simulation-based inference}

\author{Chlo\'{e} Padois$^1$\orcidlink{0009-0001-1380-9488}, Friedrich Anders$^1$\orcidlink{0000-0003-4524-9363}, Daniel del Ser$^{2,1}$\orcidlink{0000-0001-6776-3211}}
\affiliation{$^1$Departament de Física Quàntica i Astrofísica, Institut de Ciències del Cosmos (ICCUB), Universitat de Barcelona (IEEC-UB), Martí i Franquès 1, 08028 Barcelona, Spain}
\affiliation{$^2$Observatori Fabra, Reial Acadèmia de Ci\`{e}ncies i Arts de Barcelona (RACAB), Rambla dels Estudis, 115, E-08002 Barcelona, Spain}

\begin{abstract}
In the context of the rapidly growing field of galactic exoplanet population studies, we present an exoplanet simulator capable of generating synthetic populations for millions of stars in a few seconds \citep{Padois2025}. Our framework combines observed exoplanet detections with the latest planetary formation models, incorporating dependencies between host star properties and planet occurrence rates, orbital period distributions, and other key demographic features. However, many model parameters remain uncertain and model-dependent.
To constrain some of the most critical ones, we couple our simulator with simulation-based inference (SBI): we define a set of 44 free parameters describing the population of different planet types (occurrence rates, mass-period distribution, mass-radius relation, etc.), run millions of simulations sampling from broad prior distributions, simulate their detectability by {\it Kepler}, and train a machine-learning algorithm to learn the relationships between input parameters and the resulting “observed” synthetic populations. The best-fit parameters are then inferred by comparing model outputs to real observations, for which we adopt the {\it Kepler} confirmed planet catalogue as our reference dataset, since it provides the largest homogeneous sample from a single facility. We present preliminary results for the planet occurrence rate as a function of stellar mass and metallicity, across different planet types, along with constraints on several key simulation parameters such as the orbital period distribution and the inclination dispersion within planetary systems. These refined parameters enable us to refine our yield predictions for upcoming exoplanet detection missions, including PLATO, {\it Roman}, and HAYDN.
\end{abstract}


\maketitle

\section{Introduction}

Despite the rapid growth of the number of detected exoplanets in the past decades, our knowledge of exoplanet formation and evolution is limited by observational biases and detection performance. Most confirmed exoplanets are situated in the Solar Neighbourhood (80\% of them at less than 1 kpc from the Sun), but the micro-lensing technique already allowed the detection of around 200 confirmed planets at distances between 5 and 8 kpc, opening the way to the study of the exoplanet population from a Galactic perspective. 
Since the introduction of the Galactic Habitable Zone (GHZ) concept by \cite{Gonzalez2001}, the question of how planets are distributed in the Milky Way has recently regained significant interest (e.g. \citealt{Baba2023, Boettner2024, Padois2025, Spitoni2025}).

From this Galactic point of view, we can ask how homogeneous the exoplanet population is at Galactic distance scales, what shapes the exact relations between host star properties and exoplanet formation, or how the planet population is influenced by Galactic evolution (e.g. chemistry and radial migration). To address these questions, we need to simulate observations of synthetic exoplanet populations and compare the results with the actual observations from planet-detection missions.

Different flavours of planet population synthesis (PPS) models exist \citep{Mordasini2015}. Most of them simulate the exoplanet formation from the protoplanetary disc via pebble or planetesimal accretion (see \cite{Burn2024} for a review). Nevertheless, applying any of those detailed PPS models to a Galactic-scale sample would be computationally very expensive. We therefore developed a new framework to directly simulate the final exoplanet population in its stable state, combining traditional PPS predictions with constraints from earlier observations.

In Sect. \ref{sec:v1_code} we summarized the first version of the code, before presenting how we refined the model using simulation-based inference (SBI) in Sect. \ref{sec:sbi}. In Sect. \ref{sec:results} we show the preliminary results we obtain with the best predicted values, and we conclude in Sect. \ref{sec:concl}.

\section{Our previous attempt at simulating the Galactic exoplanet population}\label{sec:v1_code}

In \cite{Padois2025}, we presented a first version ({\tt v1}) of our fast exoplanet population synthesis modelling. It can be summarized in two steps: (1) For each star, depending on its mass and metallicity, we determine how many planets of different types it hosts. (2) To each planet, depending on its category (Earth-like, Super-Earth, Neptune or giant), we assign physical parameters like orbital period, inclination, radius, mass, etc. (based on prescriptions inspired by the literature). For example, the occurrence rates were based on the works of \citet{Burn2021} and \citet{Narang2018}.

\begin{figure*}
  \centerline{\vbox to 0pc{\hbox to 10pc{}}}
  \includegraphics[width=\textwidth]{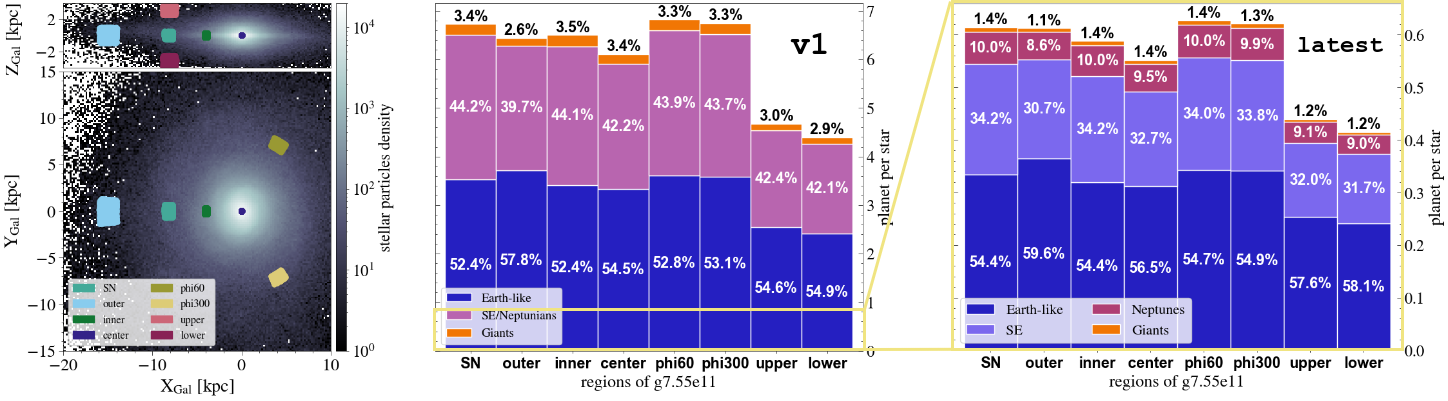}
  \caption{One of the main results from \citet{Padois2025}: the variation of the frequency of planet types as a function of position in the Galaxy. Left: definition of Galactic regions. Middle: Number of planets per star for each region and planet type, obtained with the first version of our code. Right: same as middle panel but with our latest version (note the different scales in the middle and right panels).}
  \label{fig:padois2025_summary}
\end{figure*}

We applied this \texttt{v1} of our exoplanet simulation to different regions of a Milky-Way analogue galaxy simulation (\texttt{g7.55e11} from the NIHAO-UHD suite, \cite{Buck2020}; see Fig. \ref{fig:padois2025_summary}). The main results are that we predict that Earth-likes are most likely the dominant type of planets in all Galactic regions, even if they remain significantly under-represented in current observations, and that we confirm that metal-poor regions likely host fewer exoplanets per star on average, compared to metal-richer regions \citep[as also predicted by][]{Boettner2024}.

To evaluate the accuracy of our framework, we then applied our exoplanet simulator to the stars in the \textit{Kepler} target list. For all simulated planets, we simulated if their transit would be detectable by \textit{Kepler}, based on their radius, orbital period, host star radius and magnitude, reproducing the most important of \textit{Kepler}'s sensitivity features. While our simulation reproduced some of the features seen in the {\it Kepler} data, a detailed comparison revealed that our model over-predicted the number of detected planets in the \textit{Kepler} field by a factor of $\sim10$ and also did not match the observed distribution in the period-radius diagram very well.
We already speculated in Paper I that the significant over-production of exoplanets is likely mostly coming from our occurrence rate model, based on NGPPS\footnote{New Generation Planetary Population Synthesis} output, \citet{Emsenhuber2025}, and clearly shows that we need to adjust our model.

\section{Improving our exoplanet population synthesis with SBI}\label{sec:sbi}

In order to obtain a more realistic simulation, we made major architecture changes in the code, identifying 44 key parameters which need to be constrained. Those parameters can be divided in 4 groups, depending on what they influence. See Table \ref{tab:free_params_SBI} for a summary.
The major changes with respect to \texttt{v1} include a) splitting the Super-Earths/Neptunes category in two different groups; b) changing the occurrence rate modelling: in \texttt{v1} we combined and interpolated literature references \citep{Narang2018, Burn2021}, which we replaced in the latest version by the use of adjustable functions (Johnson's $S_U$ distributions); and c) allowing for updates to many of the involved parameters.

\begin{table}[h!]
    \centering
    \begin{tabular}{l c c} \hline\hline
        parameter category              & total number & well constrained \\ \hline 
        occurrence rate \& multiplicity & 19    & 8 \\
        period/mass distribution        & 18    & 14 \\
        mass-radius conversion          & 6     & 4 \\
        inclination                     & 1     & 1 \\ \hline
    \end{tabular}
    \caption{Free parameters in the latest version of our model, adjusted with SBI using {\it Kepler} data, grouped by their influence on the exoplanet simulation process. We specify in the right column for how many parameters we obtain good constraint for a training on 3 million simulations.}
    \label{tab:free_params_SBI}
\end{table}

As we work in a high-dimensional parameter space, it is delicate to model the correlated influence of all the model parameters. We use SBI to better constrain 44 parameters (see Table \ref{tab:free_params_SBI}) and study their correlations. Our SBI methodology, summarized in Fig. \ref{fig:padois2026_summary}, consists of the following steps:

\begin{figure*}[h]
  \centerline{\vbox to 3pc{\hbox to 1pc{}}}
  \includegraphics[width=\textwidth]{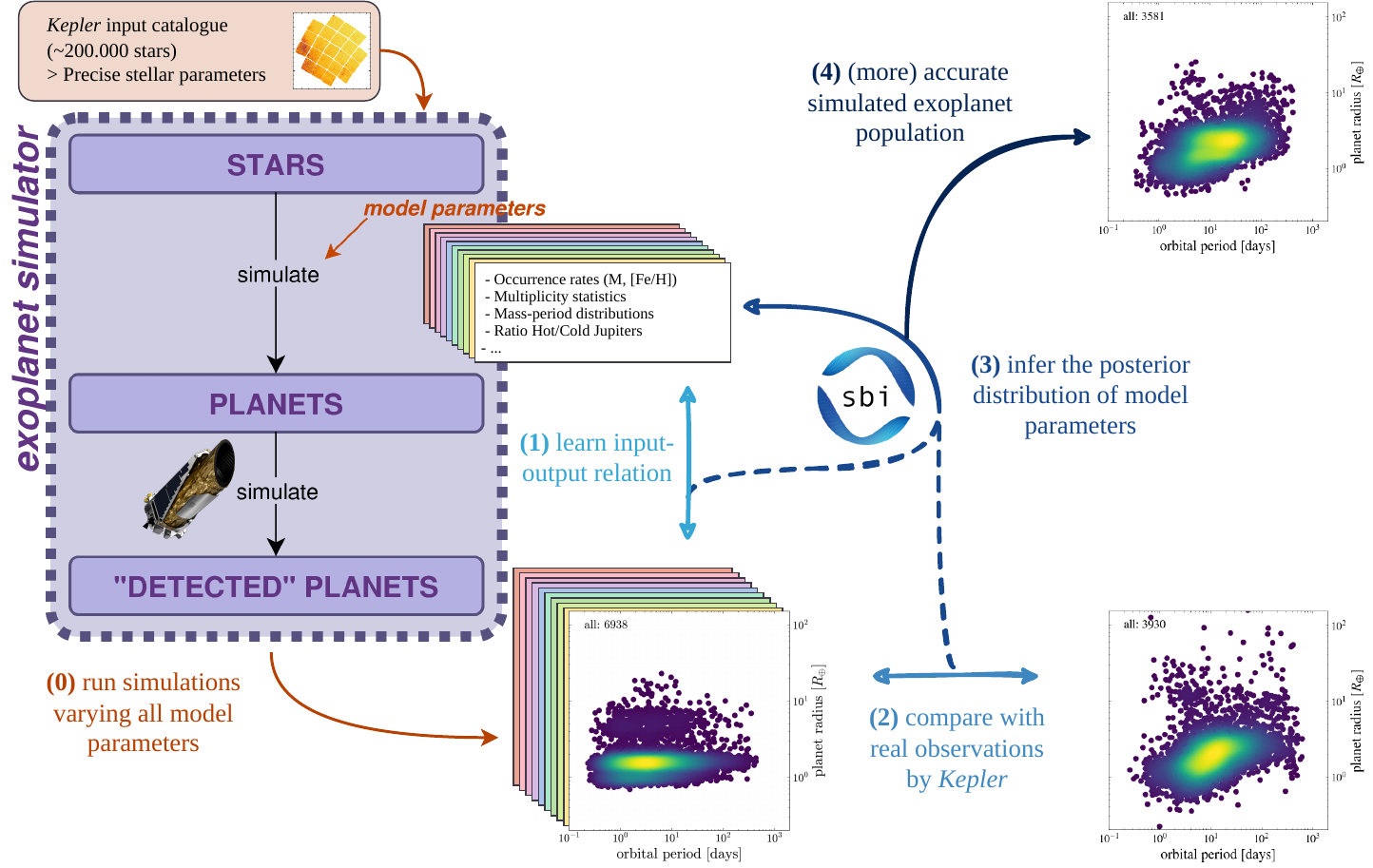}
  \caption{Visual summary of the simulation-based inference method used to optimise our exoplanet population synthesis model (Padois et al., in prep.).
  }
  \label{fig:padois2026_summary}
\end{figure*}

\begin{enumerate}
    \setcounter{enumi}{-1}
    \item Randomly varying all model parameters to cover the full parameter space, we run millions of forward PPS simulations.
    \item Core of the SBI process: we train a neural network to learn the relationship between the sampled input model parameters and the associated output, in our case the statistical properties of the \textit{Kepler}-``detectable'' exoplanet population.
    \item We define our reference observation: the candidate exoplanets actually detected by \textit{Kepler}. We use the sources classified as candidates by the \textit{Kepler} pipeline (source with ``\texttt{CANDIDATE}'' flag in Q1-Q17 DR25 table\footnote{\url{https://exoplanetarchive.ipac.caltech.edu/cgi-bin/TblView/nph-tblView?app=ExoTbls&config=q1_q17_dr25_koi}}), guaranteeing an homogeneous detection process \citep{Thompson2018}.
    \item Given the reference observation and the trained neural network, we infer the posterior distribution for all 44 free parameters to best reproduce \textit{Kepler} observations.
    \item Finally, we run the exoplanet simulation using the most probable value(s) from the posterior distribution for each parameter and update our fiducial PPS model from {\tt v1} to {\tt latest} (see Figs. \ref{fig:padois2025_summary} and \ref{fig:kepler_vs_simu}).
\end{enumerate}

\section{Preliminary results}\label{sec:results}

\begin{figure}[h]
  \centerline{\vbox to 0pc{\hbox to 1pc{}}}
  \includegraphics[width=\textwidth]{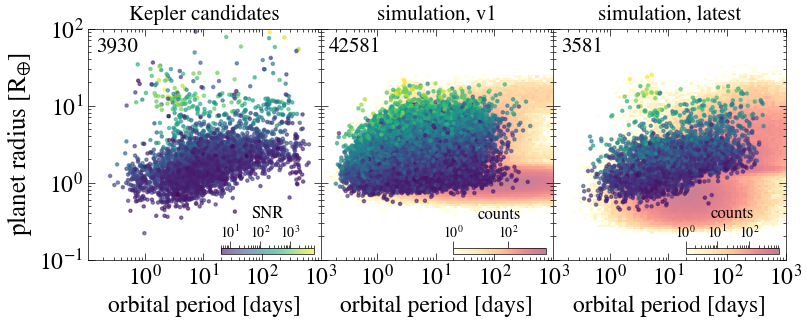}
  \caption{Planets in the \textit{Kepler} field of view. Left panel: \textit{Kepler} candidates from DR25, used as reference observation. Middle panel: simulated exoplanets with the first version of our model \citep{Padois2025}. Right panel: simulated exoplanet with our best version of the simulator, after constraining the parameters with SBI. All populations are coloured by signal-to-noise ratio (SNR) value (common colour scale), and for both simulated population we show in background the total population, before applying any detectability condition.}
  \label{fig:kepler_vs_simu}
\end{figure}

We run our exoplanet simulation, using for each parameter the most probable value (maximum of the posterior distribution). We obtain a realistic distribution in radius-period space, as shown in the right panel of Fig. \ref{fig:kepler_vs_simu}.
The main improvement with respect to the previous version perhaps is the predicted number of detected exoplanets: while our first version (tied closely to the predictions of the NGPPS model) was producing more than ten time too much detectable planets compared to \textit{Kepler} candidates, with the latest version we now obtain a similar number to the observation. Drawing 100 random samples from the posterior distribution, we obtain $3\,819 \pm 212$ detectable exoplanets, while 3\,930 candidates exoplanets have been identified in \textit{Kepler} DR25.

Applying our latest model to the same Galactic regions as presented in Sect. \ref{sec:v1_code}, we obtain similar variation between the different regions, now considering Super-Earths and Neptunes as two distinct categories. The results are shown in the rightmost panel of Fig. \ref{fig:padois2025_summary}. The main change is in the total number of planets per star, diminishing by a factor of $\sim10$ with respect to \texttt{v1}, which is a direct consequence of the improved adjustment of the occurrence rate and planet multiplicity as a function of stellar mass and metallicity. The derived occurrence rate functions, as well as a deeper analysis of all constraints obtained on the model parameters will be discussed in Padois et al. (2026, in prep.).

\section{Conclusions and outlook}\label{sec:concl}

\begin{figure}
\centering
  \centerline{\vbox to 1pc{\hbox to 10pc{}}}
  \includegraphics[width=0.6\textwidth]{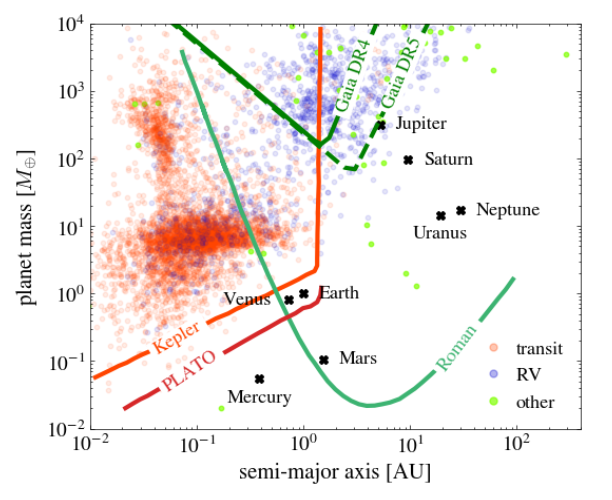}
  \caption{Estimated sensitivity threshold of different missions: \textit{Kepler}, \textit{Roman}, \textit{Gaia} DR4 \& DR5, and PLATO. Confirmed exoplanets are coloured by detection method and the Solar System planets are shown in black for reference.
}
  \label{fig:sensibility_plot}
\end{figure}

We have presented preliminary results from an upcoming paper that pioneers the use of simulation-based inference for exoplanet population studies. We were able to significantly improve the performance of our exoplanet population simulator, yielding an overall population that matches the {\it Kepler} exoplanet census in many key aspects (Fig. \ref{fig:kepler_vs_simu}). 

Now that we are more confident in the ability of our simulator to generate a realistic population, we can also use it to make more solid predictions for other exoplanet detection missions. For example, we ran preliminary tests for the PLATO southern field (LOPS2; \citealt{Nascimbeni2025}): we obtain in mean $4168\pm 84$ detectable planets for an observation period of two years, which is coherent with latest estimate by \cite{Cabrera2026}.

The future data releases from \textit{Gaia} \citep{GaiaCollaboration2016} and PLATO \citep{Rauer2025} will open a new field of possibility, giving us access to homogeneously-detected samples in under-explored regions of the radius-period space. As shown in Fig. \ref{fig:sensibility_plot}, \textit{Gaia} DR4 and DR5 will be sensitive to long-period giants, while PLATO is expected to detect small rocky planets, potentially in their host star habitable zone. In the coming years, thanks to those new datasets, we will be able to further improve our model, by obtaining estimations for parameters currently poorly constrained due to a lack of statistically significant number of detections.

\begin{acknowledgements}
This work is funded by the Horizon Europe Marie Skłodowska-Curie Actions Doctoral Network MWGaiaDN (Grant agreement No. 101072454; \url{https://www.mwgaiadn.eu/}). 
This work was partially funded by the Spanish MICIN/AEI/10.13039/501100011033 and by the `ERDF A way of making Europe' funds by the European Union through grant RTI2018-095076-B-C21 and PID2021-122842OB-C21, and the Institute of Cosmos Sciences University of Barcelona (ICCUB, Unidad de Excelencia `Mar\'{\i}a de Maeztu') through grant CEX2019-000918-M. FA acknowledges financial support from MCIN/AEI/10.13039/501100011033 through a RYC2021-031638-I grant co-funded by the European Union NextGenerationEU/PRTR.\end{acknowledgements}

\bibliographystyle{aa}
\bibliography{Sample}

\end{document}